# Nonlinear Analysis of Experimental Noisy Time Series in Fluidized Bed Systems

E. CAPUTO, S. SELLO and V. MARCOLONGO[*]

CISE - Tecnologie Innovative, Via Reggio Emilia 39, 20090 Segrate (Milano), Italy

**Abstract** - This paper describes the application of some numerical techniques to analyze and to characterize the observed dynamic behaviour of fluidized bed systems. Based on standard spectral and nonlinear analysis techniques of some selected time series data, we were able to point out the essential dynamical features of the observed behaviour and to characterize the complex irregular motion of the related nonlinear system. The preliminary results showed clearly that the dynamics of the considered process can be nonrecurrent and governed by chaotic deterministic rules rather than by stochastic ones. Moreover, the computed embedding and correlation dimensions confirmed the reliability and usefulness of the mathematical quantities from chaos theory to determine differences in the dynamical behaviours, and to extend the classical set of dynamical similarity criteria for different scaled fluidized beds. This significant conclusion was allowed by the application of a proper filtering procedure which was able to reduce the unwanted influence of significant broadband noise detected in the sampled experimental data.

## INTRODUCTION

Nonlinear phenomena behaviours have suggested that many complex dissipative fluid dynamic processes behave randomly only in appearance. It has become evident in recent years that nonlinear systems exhibiting deterministic chaotic motions are able to generate time series with a broadband power spectra. For this reason analysis of time series from real dynamical systems represents an important current area of the research in numerous different fields of science and engineering.

[*]Current address: Vela s.r.l., Viale Teodorico 2, 20149 Milano, Italy

Recent preliminary experimental analyses and model simulations have clearly indicated that the multi-phase flow in fluidized bed systems can be described as a chaotic deterministic process (Schouten and Van den Bleek [1,2], Daw [3,4]). In particular, it is suggested that the chaotic time-dependence characteristics of these systems should be included in the scaling criteria. For example, the application of dimensionless similarity groups in fluidized bed scaling is based on the non rigorous assumption that a fluidized bed system is an ordered dissipative predictable process. Nevertheless, the experimental evidence of the existence of a generic chaotic behaviour forces to extend the above criteria with different characteristic quantities from chaos theory, such as embedding and correlation dimensions, or the Lyapunov exponents [5]. The dependence of the numerical values of these mathematical quantities on fluidization and on geometrical parameters, could be an interesting research topic for a better insight of the fundamental rules governing the fluidized bed dynamics.

Moreover, considering that the hydrodynamics of fluidization is an intrinsic non-linear phenomenon, described by low-dimensional attractors, it results very useful an experimental characterization of the dynamical behaviour through the non-linear time series analysis performed on measured time series of some significant physical variable, such as the void fraction or the pressure fluctuations. In this framework the main task is to experimentally determine the existing differences in the dynamical behaviours as a function of varying operating conditions, and then to characterize or classify the observed different fluidization regimes, besides establishing, in a quantitative way, the dynamical similarities between real set-up and scaled fluidized beds.

The aim of this paper is to give a further contribution, in the engineering applications of fluidized bed systems, to strengthen the existing connections and the meaningfulness of the nonlinear and chaotic approach, for a more reliable and complete description and classification of the real physical processes involved. In the nonlinear analysis of the experimental time series, we stress the importance of using some proper specific numerical techniques, designed in order to correctly handle the original data, especially when there exists a high level broadband noise contamination.

In the first section we describe the experimental set up and the scaling criteria followed. Besides, we give information about the measurements taken in order to derive scalar time series for the nonlinear dynamical analysis.

In the second section we briefly report the experimental time series analysis tools and some of the current techniques designed to capture the dynamical features of the related physical systems. We mention, in particular, several general problems associated with reliable numerical estimates of dynamical quantities, such as correlation dimensions, and possible methods to overcome these difficulties.

In the third section we describe the principal results from the nonlinear analysis of some selected time series both of differential pressure and force fluctuations. Furthermore, in the last section, we discuss briefly about the information derived from the experimental analysis besides possible connections with a model simulation.

## EXPERIMENTAL SET-UP AND MEASUREMENTS

The cold model used had a circular section with a 300 mm inner diameter, a 1.5 m total height. The air distributor was made of 56 tuyeres with 404 horizontal orifices totally (Fig. 1), the orifices diameter was 2.0 mm and their axes was 46 mm from the bottom plate. The 40 mm thick walls were made of Plexiglas®.

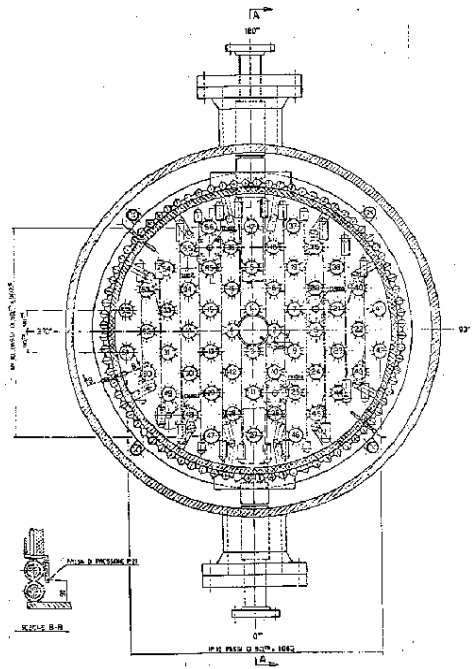

Fig. 1. Fluidizing air distributor plant

The scaling rules [6,7] adopted led to the following relations:

$$\frac{U_C}{U_H} = \left(\frac{\mu_C}{\mu_H} * \frac{\varrho_{H_f}}{\varrho_{C_f}}\right)^{\frac{1}{3}} \tag{1}$$

$$\frac{D_C}{D_H} = \left(\frac{\mu_C}{\mu_H} * \frac{\varrho_{H_f}}{\varrho_{C_f}}\right)^{\frac{2}{3}} \tag{2}$$

$$\frac{\varrho_{C_f}}{\varrho_{H_f}} = \frac{\varrho_{C_s}}{\varrho_{H_s}} \tag{3}$$

$$\frac{D_C}{D_H} = \frac{d_C}{d_H}, \frac{H_C}{H_H} = \frac{d_C}{d_H} \tag{4}$$

U indicates the superficial gas velocity, µ the dynamic viscosity, ρ the density, D the bed diameter, d the particle diameter, H the bed static height; the indexes meaning is C for cold model, f for fluid, H (Hot) for combustor, mf for minimum fluidization, p for particle.

The first two relations are equivalent to the Reynolds, based on the mean particle diameter and fluid properties, and Froude's numbers coincidence between the cold model and the reference combustor, which is a 1 MWt (at atmospheric pressure) prototype owned by ENEL (Italian National Electric Power Board).

The choice was to use, in the cold model, air at ambient conditions so it followed from the preceding relations a geometric scale equal to 1/4, a kinematic scale equal to 1/2 (it must be the square root of the geometric scale) and a solid to fluid density scale equal to 1.

In Table 1 the combustor reference conditions are reported at the time the cold model had to be designed. A large amount of bronze powder for the cold model was sieved yielding to a 183 µm mean particle diameter and a very similar scaled particle size distribution.

Unfortunately, by the time the cold model had been installed and most of the experiments were conducted, the combustor was functioning under the conditions reported in Table 1, third column. So all of the bubble frequency measurements taken at the fluidized bed combustor were executed under the changed conditions. The major change was the sand mean diameter, from 716 µm to 677 µm; a little variation in the particle mean diameter changes very much the Archimedes number (it can replace the Froude number in the similitude check). The only possibility to modify the cold model conditions towards a better similitude configuration, in a short time, was to raise the temperature to the highest possible limit of 40 °C. In spite of the temperature increase the cold model could not be performed at ideal similitude conditions as indicated in Table 1 last two columns.

Table 1. - Combustor and cold model significant set of parameters.

|  | COMBUSTOR *(reference conditions)* | COLD MODEL *(design)* | COMBUSTOR *(actual)* | COLD MODEL *(actual)* |
|---|---|---|---|---|
| P *at bed surface* (kPa) | 117.7 | 101.3 | 103 | 102 |
| T (°C) | 850 | 26 | 850 | 40 |
| $\rho_f$ (kg/m$^3$) | 0.365 | 1.1795 | 0.319 | 1.135 |
| $\mu_f$ ($10^{-5}$ Pa*s) | 4.6 | 1.839 | 4.6 | 1.906 |
| $\nu_f$ ($10^{-5}$ m$^2$/s) | 12.6 | 1.56 | 14.4 | 1.680 |
| $\rho_s$ (kg/m$^3$) | 2600 | 8404 | 2600 | 8794 |
| D (m) | 1.18 | 0.293 | 1.18 | 0.300 |
| $H_{st}$ (m) | 1.0 | 0.25 | 1.0 | 0.25 |
| $d_p$ (μm) | 716 | 179 | 677 | 183 |
| $U_{mf}$ (m/s) | 0.22 | 0.11 | 0.20 | 0.11 |
| $Re_{p,mf}$ | 1.3 | 1.3 | 0.9 | 1.2 |
| $Re_{D,mf}$ | 2060 | 2067 | 1636 | 1964 |
| Ar | 1614 | 1648 | 1194 | 1650 |

The physical variables measured and analyzed were the differential pressure on a vertical direction between two points inside the fluidized bed and the dynamic force (only for the cold model) at the edge of a test tube (outer diameter 8 mm) immersed within the bed and supported at each end by two strain gauge force transducers, an horizontal and a vertical one.

The gap between the two pressure measurements points was 5 mm in the cold model and 20 mm in the combustor. The measurements were executed using differential pressure probes connected to a differential pressure transmitter (mod. Valydine P432D, 1380 Pa full scale). The probes were basically made of two parallel small steel tubes connected through plastic tubes to the transmitter. The height at which the probes were placed was 807 mm from the combustor bottom plate and 204 mm from the cold model bottom plate. It was planned to place the probe tips at four different radial position, i.e. bed centre, one fourth, half and three fourths of the radius from the bed centre, but inside the combustor the probe tip coverings were damaged before any useful measurement could be taken when the probe was at one fourth and at half of the radius from the centre. So the only time series data available from the combustor refer to only two radial positions.

The differential pressure transmitter signal was recorded by a dynamic signal analyzer (mod. ONO-SOKKI CF-360) at 256 Hz sample rate using an anti-aliasing filter having an attenuation of 80 dB at 61% the sample frequency, so the range investigated was 100 Hz. The typical time record was 124 seconds long (31744 data). Every 1024 data the analyzer performed a FFT, so Power Spectral Density (PSD) function averages based on 31 blocks were obtainable with a frequency resolution of 0.25 Hz.

The force transducer signal was recorded by the same analyzer but at a sample rate of 128 Hz, so the time record was 248 s long and the frequency resolution of the PSD was 0.125 Hz.

## DATA ANALYSIS

Analysis of time series from experimental systems is an important issue to characterize complex real behaviour and to indicate some possible ways to correctly model the related physical phenomena. On the other hand, deterministic chaos and fractal structure in dissipative fluid dynamic systems are now recognized among the principal common nonlinear features [8]. The most used tools for these analysis are based on spectral methods, i.e. Fourier transforms (and related quantities), of the data. However, spectral analysis lies essentially on the framework of linear systems and then it is not sufficiently adequate to handle intrinsic nonlinear behaviour. When we pass from discrete lines to a broadband power spectrum, the common statement is that the dynamics was changed from a periodic or recurrent behaviour to an irregular, random like motion, typical of external noise due to the measurement devices. On the other hand, recent studies have shown in an impressive way that complex nonlinear dissipative systems can exhibit low dimensional deterministic chaos which can generate a time series whose spectrum is broadband or continuous. The distinction between chaotic behaviour and noise is then possible through the characterization of dynamic evolution of the system: in fact only deterministic systems can evolve in time nonperiodically and with impredictability on finite strange attractors that live in a proper phase space of finite dimension. On the other side, random like behaviour, or noise, cannot evolve on finite geometrical sets, but tend to occupy an arbitrarily large portion of space available. After observing an apparently random experimental signal, it would be very opportune to possess the ability of distinction between deterministic chaotic (rather than multiperiodic or quasiperiodic) and stochastic or random behaviour. Furthermore, once recognized a chaotic phenomenon it is useful to give some quantitative characterization. The important physical issue, thus, is whether the observed erratic behaviour can be modeled and described by a few nonlinearly interacting dependent variables or it is more adequately represented by an infinite degrees-of-freedom stochastic model.

As said, standard techniques of experimental data analysis are based on Fourier methods; however such tools, in the reign of nonlinear dynamics, can be qualitative only, and can be justified a posteriori. In some cases these methods can furnish some useful preliminary information before trying to compute the quantitative characteristics of the observed motions, e.g. dimensions, Lyapunov's spectrum, entropies, etc. The falloff of power spectra at high frequencies indeed, can result in a possible mean of distinguishing deterministic chaos from stochastic motion, as well described in the work by Sigeti and Horsthemke (1987)[9].

Besides these qualitative techniques, there are now many powerful tools of nonlinear dynamics analysis: for example, correlation dimension, mainly due to Grassberger and Procaccia (1983), has become the most popular and widely used measure of complexity on experimental strange attractors [10]. Typical experimental situations concern only a single scalar time series; while the related physical systems possess many essential degrees of freedom. The powerfulness of correlation dimension like methods, rely on the reconstruction of the whole system's trajectory in an *"embedding space"* using the method of delay-time. The reliability of computations, performed on the reconstructed trajectory, is guaranteed by a notable theorem by Takens and Mañé (1981) [11].

Let a continuous scalar signal x(t) be measured at discrete time intervals, $T_s$, to yield a single scalar time series:

$$\{ x(t_0), x(t_0+T_s), x(t_0+2T_s),..., x(t_0+NT_s) \}.$$

We assume x(t) to be one of n possible state variables which completely describe our dynamical process. For practical applications, n is unknown and x(t) is the only measured information about the system. We suppose, however, that the real trajectory lies on a d-dimensional attractor in its phase space, where: d≤n. Packard et alt. and Takens have shown that starting from the above time series it is possible to *"embed"* or reconstruct a *"pseudo-trajectory"* in an m-dimensional embedding space through the vectors (embedding vectors):

$$\begin{aligned}
\underline{y}_1 &= (\ x(t_0),\ x(t_0+\tau),..., x(t_0+(m-1)\tau)\ )^T \\
\underline{y}_2 &= (\ x(t_0+l),\ x(t_0+l+\tau),..., x(t_0+l+(m-1)\tau)\ )^T \\
&\ \ \ ... \\
\underline{y}_s &= (\ x(t_0+(s-1)l),\ x(t_0+(s-1)l+\tau),..., x(t_0+(s-1)l+(m-1)\tau)\ )^T.
\end{aligned} \quad (5)$$

Here τ is called *"delay-time"* or lag, and l is the sampling interval between the first components of adjacent vectors [12]. There is an essential quantity to consider in the reconstruction process: the *"window length"* defined as: w=(m-1)τ, which represents the time spanned by each embedding vector. A selection of proper values of parameters in the embedding procedure is a matter of extreme importance for the reliability of results, as well pointed out in many works [13,14]. The delay time, τ, for example, is introduced because in an experiment the sampling interval is in general chosen without an accurate prior knowledge of characteristic time scales involved in the process.

Takens formal criterion tells us how embedding dimension m and attractor dimension d must be related to take a proper embedding, i.e. with equivalent topological properties:

$$m \geq 2d + 1. \tag{6}$$

Fortunately, for practical applications, this statement is generally too conservative and thus it is adequate and correct a reconstruction of attractor in a space with a lower dimensionality.

After the embedding space reconstruction we can calculate the dimension of related attractor using some algorithm. In this work we consider a proper version of the original Grassberger-Procaccia algorithm, following the lines of Theiler (1990) [15] and Martinerie et alt. (1992) [14]. In an m-dimensional embedding space the *"correlation integral"* is defined as:

$$C_m(\alpha,N,r) = \frac{2}{(N+1-\alpha)(N-\alpha)} \sum_{j=\alpha}^{N-1} \sum_{i=0}^{N-1-j} \theta(r - \|\underline{y}_i - \underline{y}_{i+j}\|). \tag{7}$$

where θ(·) is the usual Heaviside unit-step function, and α is a "residual time correlation" parameter. When α=1, we obtain the standard algorithm. This version allows a better computation of correlation integrals when there are residual time correlations in the embedding vector components.

Grassberger and Procaccia in 1983 show that the correlation dimension, $D_2$, associated to space reconstructed attractors, is given by:

$$D_2 = \lim_{m \to \infty} \lim_{r \to 0} D_2(m;r), \tag{8}$$

where $D_2$ is the extrapolated slope in the so-called *"scaling region"* of correlation integrals:

$$D_2(m;r) = \frac{d[\log(C_m)]}{d[\log(r)]}. \tag{9}$$

For practical analysis, involving experimental data, the limits in the above definitions cannot be computed, and we must take care of noise and data accuracy limitations, for small r values, and finite attractor size and data set, for larger r values. Previous works suggest that a good criterion for reliable correlation estimations, is the local constancy of window length, rather than a proper choice of separately embedding dimension and/or delay time. For this reason here we adopt the technique of an optimal constant window length, following the lines in [13]. More precisely, we start with a proper choice of delay time through the *mutual information* of Fraser and Swinney [16], considering lags that are some fraction (e.g. 1/10th) of the first local minimum. This is because a fraction of the minimum works better when experimental noise contaminated time series are to be analyzed [17]. Afterwards we determine the optimal window length following the criterion based on the broadest plateau of slopes. Finally, we compute the correlation integrals for different embedding dimensions using eqn. (7), extracting the slope information after a proper selection of scaling region, where the correlation integrals behaviour is nearly linear. It is well known that if dimension $D_2$ reaches a saturation value beyond some embedding value $\overline{m}$, the system represented by the time series is deterministic on a finite attractor, and the saturation value, $\overline{D_2}$, will be regarded as a lower bound of the number of state variables needed to describe the related dynamics. On the other hand, if there is any kind of saturation of dimension, but all the values tend to follow a rule proportional to embedding dimension, then the system behaves like a stochastic process with an arbitrarily large number of degrees of freedom.

Methods of dynamical system analysis can be strongly limited by typical features of experimental situations. Correlation dimension techniques, in particular, are based on assumptions that cannot be rigorously fulfilled by experiments, especially due to the presence of broadband noise. In real cases it can happen that the presence of noise results in a severe pitfall for correlation dimension algorithms, compromising the reliability of distinction between stochastic and deterministic behaviour. For this reason it would be of great advantage the possibility of processing experimental data in order to reduce the unwelcome noise level, without distorting the original signal. To this aim, in literature we can find many different examples of linear filters methods proposed to reduce noise on real data. Recently, however, nonlinear methods of noise reduction have been proposed which seem to work correctly when applied to real-world time series. In particular here we pay attention to a proper version recently proposed by Schreiber and Grassberger (1991)[18]. In the linear

framework, we use the following recurrent transformation:

$$\bar{x} = \sum_{i=-k}^{k} a_i \, x_{n+i} + b, \quad n=1,2,...N, \qquad (10)$$

where: $a_0 < 1$ and fixed.

Optimal values for coefficients are obtained by a proper least squares fit where the quantity: $\sum_{n=1}^{N} (\bar{x}_n - x_n)^2$, is minimized. Free parameters $a_0$ and $k$ must be defined a priori, and they depend on the level of correction and on the number of frequencies involved. Previous extensive applications on experimental fluid systems have shown the reliability and effectiveness of the above filter, especially when a broadband corrupting noise is present.

## EXPERIMENTAL RESULTS

According to the scaling principles the differential pressure main frequencies should be in a ratio of 2 to 1 between cold model and combustor and the Probability Density Function (PDF) should be identical. Both expectations were fulfilled for the bed centre fluid-dynamics as revealed by the time series L011200 (combustor at 0.88 m/s of superficial gas velocity, see Table 2) and BFRN02 (model at 0.45 m/s of $U_f$) (Figs. 2-4), while close to the wall (Figs. 5-7) the time series C011200, C021200 (combustor at 0.88 m/s of $U_f$) and BFRN01 (model at 0.45 m/s of $U_f$) PSD functions had a good similitude but the two PDFs were not identical as expected. A satisfactory similar behaviour between the two beds was realized, discrepancies from an exact similar behaviour descended probably from the difference between actual and ideal conditions the cold model should have had after changes respect to the initial combustor's reference conditions.

The non linear analysis, performed according to the above mathematical techniques, outstandingly confirmed the results evidenced through the classical Fourier analysis tools; more precisely the correlation dimension was exactly the same for series L011200 and BFRN02, i. e. 5.8 (Fig. 8), while for time series C011200 and C021200 the average correlation dimension, i. e. 6.4, was slightly different from the BFRN01 series value, i. e. 6.1.

The above fractal dimensions were obtainable only after the application of the filtering procedure described in the previous section (Eqn. (10)), since the experimental noise level detected in the original data was such that the stochastic component effects dominate in the correlation integrals calculations.

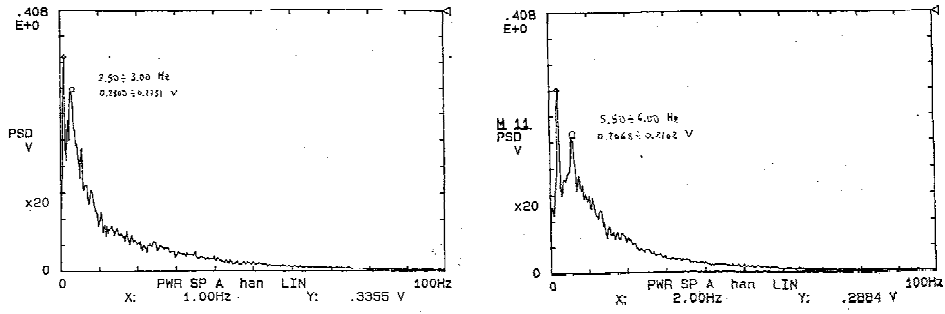

Fig. 2. ΔP PSD relative to combustor center bed (left) and of cold model (right)

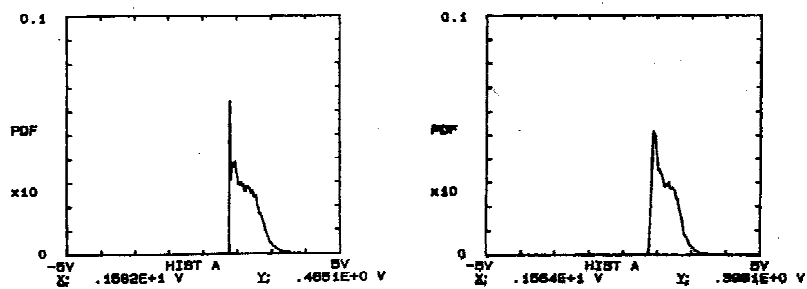

Fig. 3. ΔP PDF relative to combustor center bed (left) and of cold model (right)

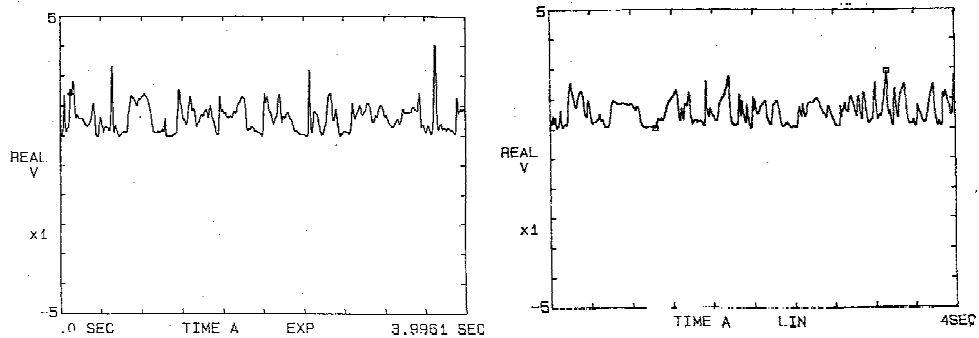

Fig. 4. ΔP signal examples relative to combustor center bed (left) and of cold model (right)

The filtering consequences are geometrically well evidenced through the attractor shapes in the reconstructed phase space, as shown in the 3-D Poincarè return maps (Fig. 9). The "sharp like" structures of the original attractor, which fade in the filtered one, is a typical noise hallmark which contaminates the data.

These preliminary outcomes confirm the substantial adequacy of the scaling criteria based on dimensionless numbers; moreover the nonlinear analysis seems, compared to the Fourier analysis, quite sensitive in detecting a situation of inperfect similarity. In fact the C011200 and C021200 series average correlation dimension was appreciably different from the BFRN01 series one, corresponding to the lack of interchangeability of the PDFs. However it must be pointed out that the whole set of non linear analysis and classical statistics and Fourier analysis calculated quantities is necessary for assessing the achievement of a complete fluidynamics similarity. For example the time series SFR033 and R081250 (see Table 2) had exactly the same embedding and correlation dimension but there was not any scaling correlation, being the dimensionless numbers very different in value ($Re_p$=19.1 and $Ar_p$=28523 for SFR033 series, $Re_p$=5.9 and $Ar_p$=2754 for R081250).

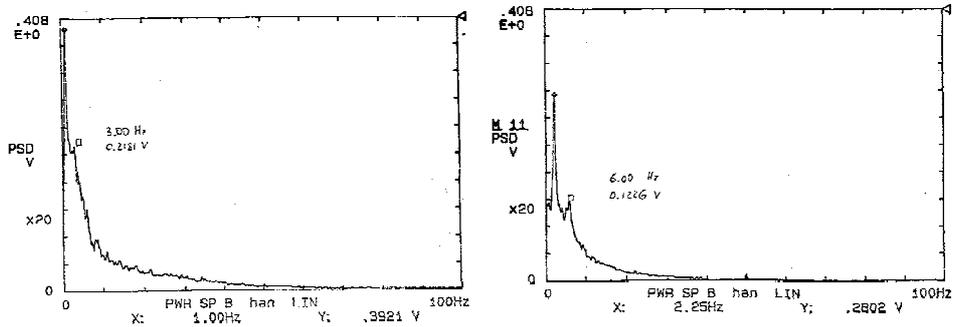

Fig. 5. ΔP PSD relative to ¾ of radius from combustor center bed (left) and cold model (right)

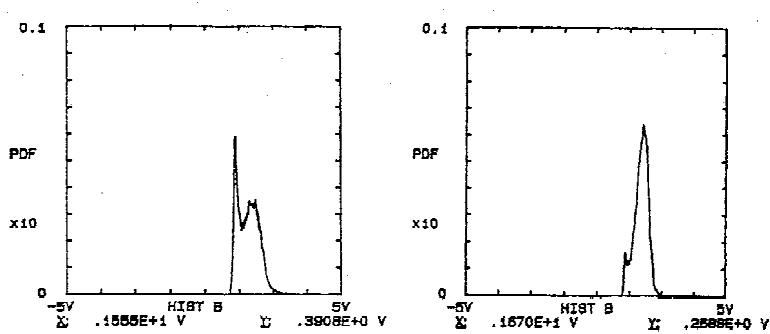

Fig. 6. ΔP PDF relative to ¾ of radius from combustor center bed (left) and cold model (right)

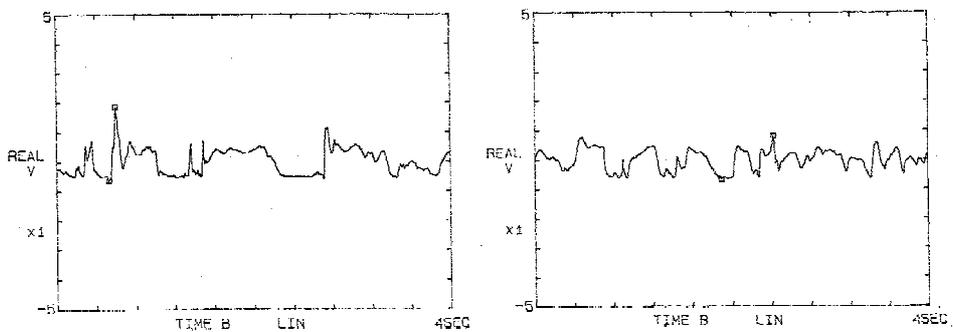

Fig. 7. ΔP signal examples relative to ¾ of radius from combustor center bed (left) and cold model (right)

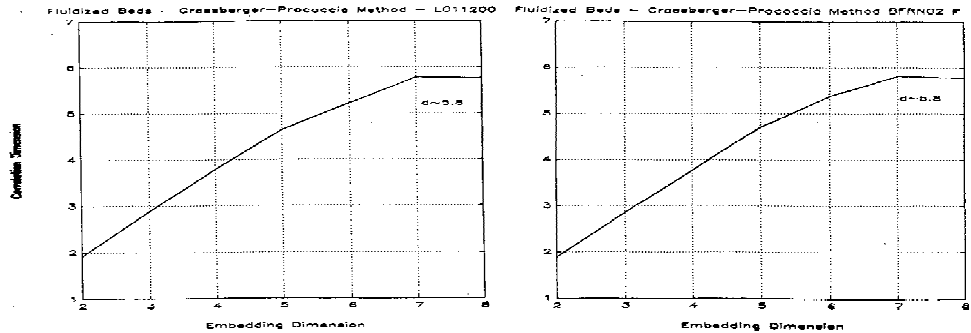

Fig. 8. Correlation Dimension vs Embedding Dimension for filtered L011200 (left) and for filtered BFRN02 (right)

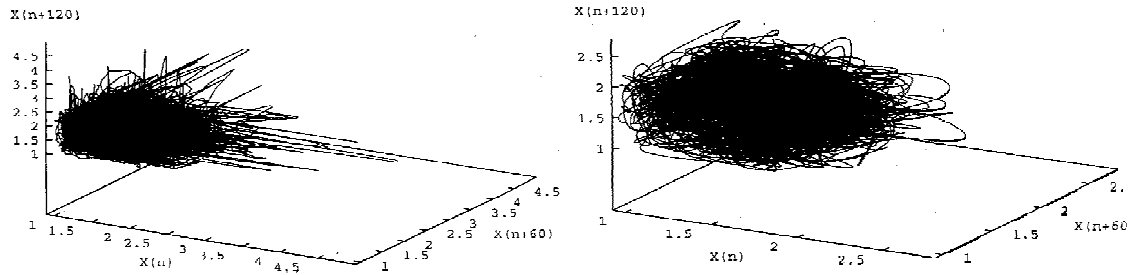

Fig. 9. 3-D Poincare' return maps for unfiltered (left) and for filtered C011200 series (right)

Part of the time series nonlinear analysis was dedicated to an introductory estimation of some physical parameter influence on the embedding and correlation dimension values. The physical parameter considered were: the fluidization superficial velocity and the radial position.

Four values of $U_f$ for the cold model at 20 °C and atmospheric pressure with bronze powder were considered, i.e. 0.13, 0.145, 0.24, 0.33 m/s; for the four time series of the differential pressure measured, i. e. BFR120, BFR109, BFR107 and BFR126 (see Table 2), correlation dimensions were respectively: 4.4, 4.45, 4.9 and 6.4, the embedding dimension was 6 for everyone but the one at the highest $U_f$. The embedding dimensionality increase is a strongest evidence than the correlation dimension of an augmented dynamical complexity.

The time series BFRN02, BFRN04, BFRN05 and BFRN01 (see table 2) of differential pressure inside the bronze powder bed at 40 °C, atmospheric pressure and 0.46 m/s of $U_f$ are relative to different radial

Table 2. Non linear analysis results.

| Series | Distance from wall | Min. Mut.Inf. | Embedding dimension | Correlation dimension | $U_f$ (m/s) | $U_f-U_{mf}$ (m/s) | $U_f/U_{mf}$ |
|---|---|---|---|---|---|---|---|
| BFRN05[2] | ½ R | 30 | 7 | ~ 6.3 | 0.47 | 0.36 | 4.27 |
| BFRN01[2] | ¼ R | 30 | 7 | ~ 6.1 | 0.46 | 0.35 | 4.18 |
| C011200[1] | ¼ R | 60 | 7 | ~ 6.3 | 0.88 | 0.71 | 5.17 |
| L011200[1] | R | 28 | 7 | ~ 5.8 | 0.88 | 0.71 | 5.17 |
| BFRN02[2] | R | 22 | 7 | ~ 5.8 | 0.46 | 0.35 | 4.18 |
| BFR126[2] | R | 13 | 7 | ~ 6.4 | 0.33 | 0.22 | 3.00 |
| BFR107[2] | R | 16 | 6 | ~ 4.9 | 0.24 | 0.13 | 2.20 |
| BFR109[2] | R | 19 | 6 | ~ 4.45 | 0.145 | 0.035 | 1.32 |
| BFR120[2] | R | 16 | 6 | ~ 4.4 | 0.13 | 0.02 | 1.22 |
| C021200[1] | ¼ R | 70 | 7 | ~ 6.5 | 0.88 | 0.71 | 5.17 |
| BFRN04[2] | ¾R | 17 | 6 | ~ 6.2 | 0.47 | 0.36 | 4.27 |
| SFR033[3] | R | 20 | 6 | ~ 4.5 | 0.43 | 0.12 | 1.39 |
| R081250[4] | R | 40 | 6 | ~ 4.5 | 0.95 | 0.68 | 3.52 |
| E2BSBDP4[2] | 70 mm | 13 | 7 | ~ 6.2 | 0.46 | 0.35 | 4.18 |
| E2BSBF4[2] | 70 mm | 22 | 7 | ~ 6.0 | 0.46 | 0.35 | 4.18 |

(1) sand in FBC at 850 °C, $U_{mf}$=0.17 m/s, $d_p \cong 677$ μm - (2) bronze powder in cold model, $U_{mf}$=0.11 m/s, $d_p$=183 μm - (3) sand in cold model, $U_{mf}$=0.31 m/s, $d_p$=677 μm - (4) quartz in FBC at 850 °C, $U_{mf}$=0.27 m/s, $d_p$=895 μm.

position, i. e. centre, one fourth, half and three fourths of radius from centre; their correlation dimensions were respectively: 5.8, 6.2, 6.3 and 6.1. It was observed that during fluidization the bulk of the bubbles would come up to the bed surface at a few centimeter from the centre, this was probably due to the lack of tuyeres at the distribution air plate centre because of the presence of the ash discharge duct (see Fig. 1).
It could be supposed that correlation dimension values reflect this peculiar dishomogeneity in bubble distribution.

A noticeable result concerns the comparison between force time series E2BSBF4 and differential pressure (measured 20 mm below the tube) time series E2BSBDP4: their correlation dimension values were very close, 6.0 and 6.2 respectively, though one variable consists of a "global dynamic information", as it is the resultant of the forces acting on an immersed tube measured at one edge, and the other variable consists of "point dynamic information", because it is the measure of the local pressure gradient. It was generally observed that the differential pressure and vertical force measured PSDs are quite similar, particularly from a certain $U_f$ on, so the nonlinear analysis results confirmed the previous evidence. In this particular case the

PSDs were not very similar (see Fig. 10), nonetheless the correlation dimensions exhibited a comparable dynamical complexity degree.

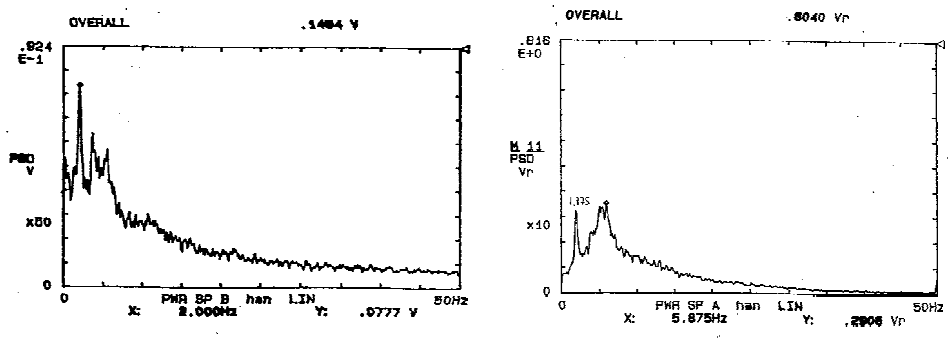

Fig. 10. E2BSBF4 (left) and E2BSBDP4 (right) series PSD

The information drawn from the experimental time series analysis may help in the mathematical formulation of proper analytical models for simulation and prediction of the dynamical behaviour. For this reason the FLUFIX/MOD.2 code (developed by ANL [19]) was used to simulate the sand bed cold model at the conditions held during the SFR033 series recording. This first comparison between an experimental time series and the calculated one was aimed at appraising the model capability of reproducing the whole dynamical complexity.

The time step used during computation was $5*10^{-5}$ s and the time interval simulated was 40 s long. The porosity (i. e. the void fraction) from the cell corresponding to bed centre and 204 mm height was the variable considered for the analysis. The sample time step was 0.001 s so the time series contained 40000 data (Fig. 11); its peak frequencies were 0.85 Hz, 2.43 Hz and 5.43 Hz.

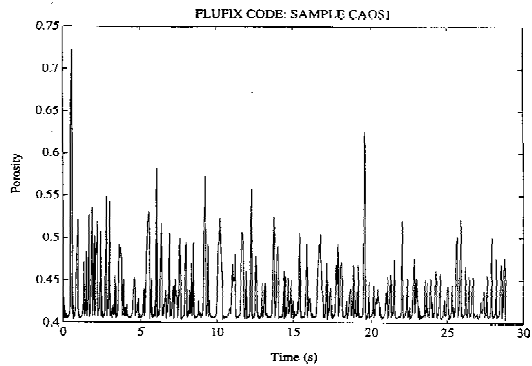
Fig. 11. Porosity computed time series with FLUFIX code

The correlation integrals computations (Fig. 12) led to a value of 2.4 for the correlation dimension and about 4 for the embedding one. These results, confirmed also by the Kolmogorov entropy estimate, $K_2$, show a substancial qualitative agreement of the model performances with the experimental data.

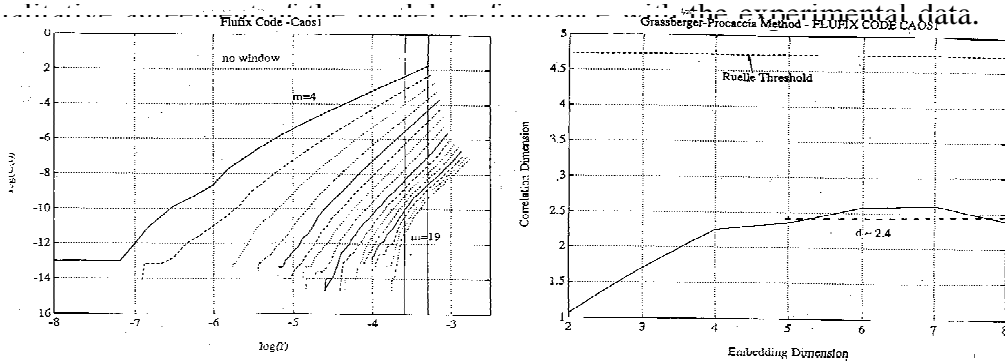
Fig. 12. Correlation integrals (left) and correlation dimensions (right) for the simulated time series with FLUFIX code

The mathematical model was a 2-D one so it could be possible that a 3-D model would have reproduced a more complex behaviour and consequently higher correlation and embedding dimension.

## CONCLUSIONS

The nonlinear analysis of some differential pressure measurements time series confirmed the reliability of certain mathematical quantities from chaos theory to point out some peculiar features of fluidized bed systems behaviour. Moreover the application of this analysis within the scaling problem allows a more complete validation of an appropriate reproduction of the fluidynamics through scaled models. Besides, our preliminary investigation suggests the possibility of the nonlinear analysis information of adding more insight to the complex real processes dynamical behaviour. To this respect further experimental and model investigations are certainly needed in order to appreciate the whole potentiality of these analyses for engineering applications.

*Acknowledgement* - This research was supported by ENEL SpA /DSR/CRT of Pisa.